\documentclass{article}
\usepackage{mn2e}
\usepackage{mc238-bib}
\usepackage{epsfig}
\begin{document}
\newcommand{\pI}{Paper~I}
\newcommand{\miHz}{\,$\mu$Hz}

\title[Velocity Analysis of PG\,1605+072]
{Time-series Spectroscopy of Pulsating sdB Stars II:
  Velocity Analysis of PG1605+072\thanks{Based on observations made
    with the Danish 1.54\,m telescope at ESO, La Silla, Chile, and on
    observations made with the Nordic Optical Telescope, operated on
    the island of La Palma jointly by Denmark, Finland, Iceland,
    Norway and Sweden in the Spanish Observatorio del Roque de Los
    Muchachos of the Instituto de Astrof\'isica de Canarias.}}

\author[S.~J.~O'Toole et al.]{S.~J.~O'Toole,$^1$ T.~R.~Bedding,$^1$ H.~Kjeldsen,$^2$
  T.~H.~Dall,$^3$ and D.~Stello$^3$\\
$^1$School of Physics, University of Sydney, NSW 2006, Australia\\
$^2$Theoretical Astrophysics Center, Aarhus University, DK-8000,
  Aarhus~C, Denmark\\
$^3$Institute of Physics and Astronomy, Aarhus University, DK-8000,
  Aarhus~C, Denmark}
\maketitle

\begin{abstract}

We present the analysis of time-resolved spectroscopy of the pulsating
sdB star PG\,1605+072. From our main observing run of 16 nights we
have detected velocity variations at 5 frequencies that correspond to
those found in photometry. Based on these data, there appears to be
change in amplitude of the dominant modes over about a year. However,
when we include extra observations to improve the frequency
resolution, we find that some of the frequencies are split into two or
three. Simulations suggest that the apparent amplitude variation can
be at least partially explained by a series of very closely spaced
frequencies around the two strongest modes. By combining observations
taken over $\sim$300 days we conclude that some of the closely spaced
modes are caused by one mode whose amplitude is varying, however this
frequency is still within $\sim$1\miHz\ of an apparently stable
frequency. Because of this kind of complexity and uncertainty we
advise caution when trying to identify oscillation modes in this star.

\end{abstract}


\section{Introduction}
\label{sec:intro}
The discovery of pulsations in hot subdwarf B (sdB) stars provides an
excellent opportunity to probe sdB interiors using asteroseismological
analysis. While it is generally accepted that sdBs are the field
analogues of Extreme Horizontal Branch (EHB) stars, many questions
remain about their formation and evolution. Subdwarf B stars (like
their EHB counterparts) are He-core burning, with 20\,000\,K\,$\la
T_{\mathrm{eff}}\la$\,40\,000\,K and 5.2\,$\la$\,log\,$g$\,$\la$\,6.2
\cite{SBK94}.

\citeone{DCru96b} showed that one possible mechanism for EHB (and sdB)
formation is strong mass loss on the Red Giant Branch. They found
that the mass-loss efficiency required to produce EHB stars does not
vary greatly with metallicity, which might explain the existence of EHB
stars in both metal-rich populations (some clusters and possibly
elliptical galaxies; \citebare{DOR95}) and metal-poor populations
\citeeg{Whit94}. Binary interaction is one possible mechanism for such
mass loss, an idea first introduced by \citeone{MNG76}.  Indeed, at
least $\sim$60\% of sub\-dwarfs appear to have main sequence companions
\cite{AWF94,J+P98}.

Pulsating sdBs (also known as EC 14026 stars)
typically have periods of 100--200\,s, photometric amplitudes
$<$\,10 mmag, and oscillate in $p$ as well as possibly $g-$modes. The
pulsations are thought to be driven by an opacity bump due to the
ionisation of heavy elements such as Fe at temperatures of
$\sim2\times10^5$K in the sdB envelope \cite{CFB97a}.

\begin{figure*}
\epsfig{file=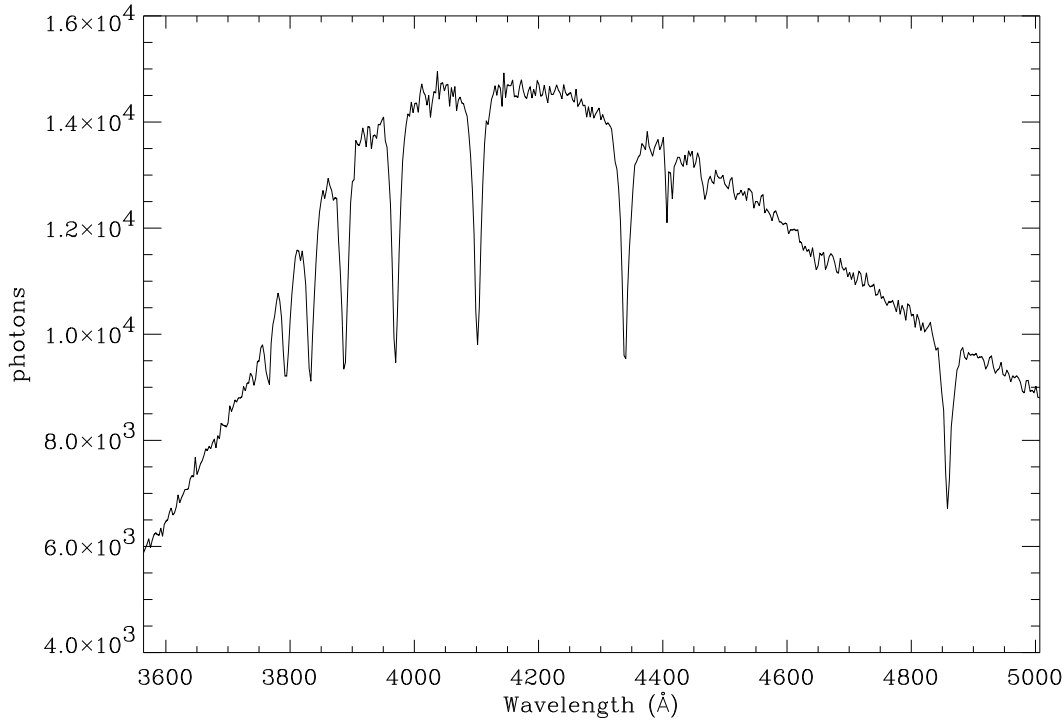,scale=0.8}
\caption{Typical spectrum of PG 1605+072 obtained at the Nordic
Optical Telescope. The Balmer sequence is visible from H$\beta$ to
H12, as well as the He \protect \textsc{i} $\lambda$4471\,\AA\ line.
The feature at $\sim$4410\,\AA\ is a bad CCD column.}
\label{fig:typispec}
\end{figure*}

Of all the known pulsating sdBs, PG 1605+072 has the most extreme
properties, with the richest power spectrum (around 50 modes), the
longest periods (up to $\sim$550 seconds), and the lowest surface
gravity (log\,$g\sim\,5.25$) \cite{ECpaperX}. The low gravity implies
that the star may have evolved off the EHB; this hypothesis is
supported by modelling by \citeone{Kawaler99}.

In \pI\ \cite{OBK00b} we reported the detection of Doppler variations
in PG 1605+072. The dominant frequencies detected corresponded to
those found in photometry \cite{ECpaperX}, and we also found evidence
of a wavelength dependence of the oscillation amplitude of the Balmer
lines. In this paper we present new radial velocity observations
covering a much longer time span. We do not confirm the wavelength
dependence, but we do see evidence for very closely spaced frequencies,
similar to the findings of \citeone{ECpaperX}. We also compare our
results with those of \citeone{WJP01}, who observed this star with
higher S/N and spectral dispersion, but over a much shorter time span
(see Section \ref{sec:comp}). We will discuss equivalent width
measurements from our spectra in a future paper.

\section{Observations}
\label{sec:obs}

The bulk of the observations were made on 11 nights over a 16-day
period in May 2000 (see Table \ref{tab:spec}), using the DFOSC
spectrograph on the Danish 1.54\,m telescope at La Silla, Chile and
the ALFOSC spectrograph mounted on the 2.56\,m Nordic Optical
Telescope on La Palma in the Canary Islands. To increase the frequency
resolution of the amplitude spectrum and allow detection of very
closely spaced oscillation modes, we also obtained observations at La Silla
for about 1 hour per night in March--April 2000, before the main run
(see Table \ref{tab:marapr-obs}). 

\begin{table}
\caption{Spectroscopic observations of PG1605+072. LS = La Silla; NOT
= Nordic Optical Telescope.}
\label{tab:spec}
\vspace{0.2cm}
\begin{center}
\begin{tabular}{lccc}
\hline 
UT-date & Telescope & No. of & No. of \\
 & & hours & spectra \\
\hline

2000  May 11 & Danish & 6.87 & 370 \\
2000  May 12 & Danish & 6.45 & 300 \\
2000  May 16--17 & NOT & 7.31 & 337 \\
2000  May 17--18 & NOT & 8.12 & 376 \\
2000  May 18 & Danish & 6.75 & 319 \\
2000  May 19 & Danish & 6.77 & 375 \\
2000  May 19--20 & NOT & 8.40 & 473 \\
2000  May 20 & Danish & 6.61 & 370 \\
2000  May 21 & Danish & 2.69 & 146 \\ 
2000  May 22 & Danish & 6.69 & 374 \\
2000  May 23 & Danish & 6.67 & 369 \\
2000  May 25 & Danish & 6.62 & 346 \\
2000  May 26 & Danish & 1.24 & \,\,70 \\
\hline 
\textbf{Total} & & 81.19 & 4225

\end{tabular}
\end{center}

\end{table}

The La Silla data (March to May) consisted of single-order spectra
projected onto a 2K LORAL CCD.\@ Pixel binning (to reduce readout
noise) and windowing (to reduce readout time) gave 66 x 500-pixel
spectra with a total wavelength range of 3650--5000\,\AA\ and a
dispersion of 1.65\,\AA\,pixel$^{-1}$. The resolution was $\sim$6\,\AA,
set by a slit width of 1.5 arcsec. The exposure time was 46\,s, with a
dead time of about 16\,s. The average number of photons per \AA\ in
each spectrum was about 1700. A similar setup was used in previous
observations (see \pI).

The NOT data were similar, with 66 x 500 binned pixels covering a
wavelength range of 3550--5000\,\AA. The dispersion was
1.65\,\AA\,pixel$^{-1}$ and the resolution was $\sim$8\,\AA, set by a
slit width of 1.0 arcsec. The exposure time was 50--70\,s, with a dead
time of only 5\,s. There was an average of about 8000 photons per \AA\
in each frame. The data from both telescopes were timestamped in
Modified Julian Date by the telescope computers to an accuracy of less
than one second.

\begin{figure}
\epsfig{file=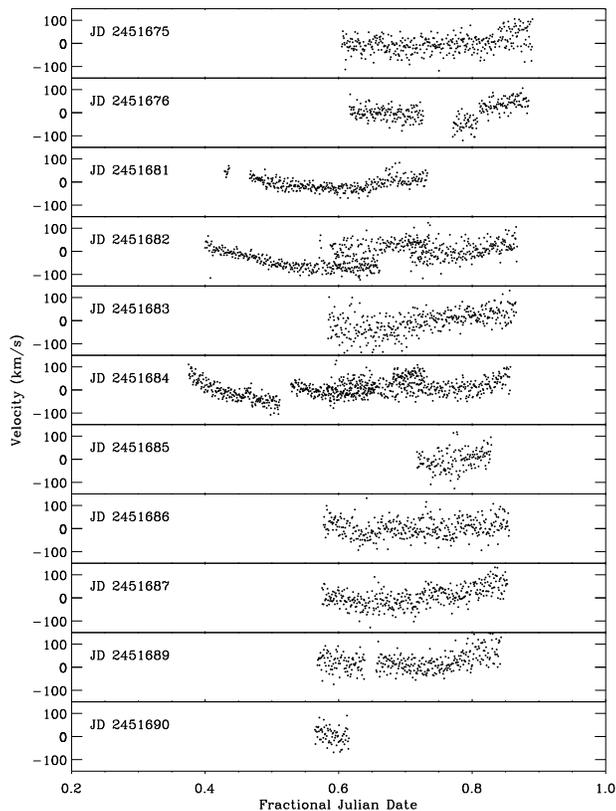,scale=0.47}
\caption{Velocity curve for H$\gamma$ from May 2000. No instrumental
variation has been corrected for. There is a 3 day gap between the 2nd
and 3rd panels from the top.}
\label{fig:velcurve}
\end{figure}

\begin{table}
\caption{Spectroscopic observations of PG1605+072 in March-April 2000.}
\vspace{0.2cm}
\begin{center}
\begin{tabular}{lcc}
\hline UT-date & No. of & No. of \\ & hours & spectra \\ \hline

2000  March 18 & 0.30 & 12 \\ 2000  March 20 & 1.43 & 77 \\ 2000 March
21 & 0.45 & 25 \\ 2000  March 22 & 0.45 & 25 \\ 2000  March 23 & 0.57
& 32 \\ 2000  March 24 & 0.65 & 37 \\ 2000  March 25 & 0.72 & 41 \\
2000  March 26 & 0.93 & 45 \\ 2000  March 27 & 0.93 & 52 \\ 2000 March
28 & 0.83 & 45 \\ 2000  March 29 & 1.17 & 59 \\ 2000  March 30 & 1.05
& 59 \\ 2000  March 31 & 1.07 & 59 \\ 2000  April 01 & 1.15 & 57 \\
2000  April 04 & 1.03 & 65 \\ 2000  April 05 & 1.42 & 77 \\ 2000 April
06 & 1.47 & 80 \\ 2000  April 07 & 1.55 & 85 \\ 2000  April 08 & 1.65
& 90 \\ 2000  April 09 & 1.57 & 86 \\ 2000  April 10 & 1.70 & 95 \\
2000  April 11 & 1.90 & 99 \\ 2000  April 12 & 1.23 & 65 \\ \hline
\textbf{Total} & 25.22 & 1367 \\
\label{tab:marapr-obs}             
             
\end{tabular}
\end{center} 
             
\end{table}

\section{Reductions}
\label{sec:red}
Bias subtraction, flat fielding and background light subtraction were
done using IRAF, and 1D spectra were extracted using an Optimal
Extraction algorithm \cite{OptEx86}. A cubic spline was fitted to the
continuum level, and the spectra were normalised to a continuum value
of unity.

The spectrum of PG1605+072 is dominated by Balmer lines (see Figure
\ref{fig:typispec}). Also present are some weak lines from He I (e.g,
$\lambda$4471 and $\lambda$4922) and He II (e.g, $\lambda$4686). The
Balmer lines are ideal to determine Doppler variations. A template
spectrum was created by averaging 20 high-quality spectra from each
night. By using a different template for each night, we are
effectively applying a high-pass filter to the data.

To avoid pixelization effects in the cross correlation analysis, all
spectra, including templates, were oversampled by a factor of 40
(using linear interpolation), and all but the Balmer line in question
(one of H$\beta$, H$\gamma$, H$\delta$, H$\epsilon$ and H8) was
cropped from each.

These spectra were modified further by subtracting 1.0 such that the
continuum level was approximately zero. The result was passed through a
half-cosine filter to smooth the transition to zero at the ends.

Each resulting template was then cross-correlated with its respective
spectra to obtain pixel displacements. The velocities were obtained by
using the known wavelengths of the Balmer lines to convert this
displacement into wavelength and then Doppler shift. See \pI\ for
further details.

\begin{figure*}
\epsfig{file=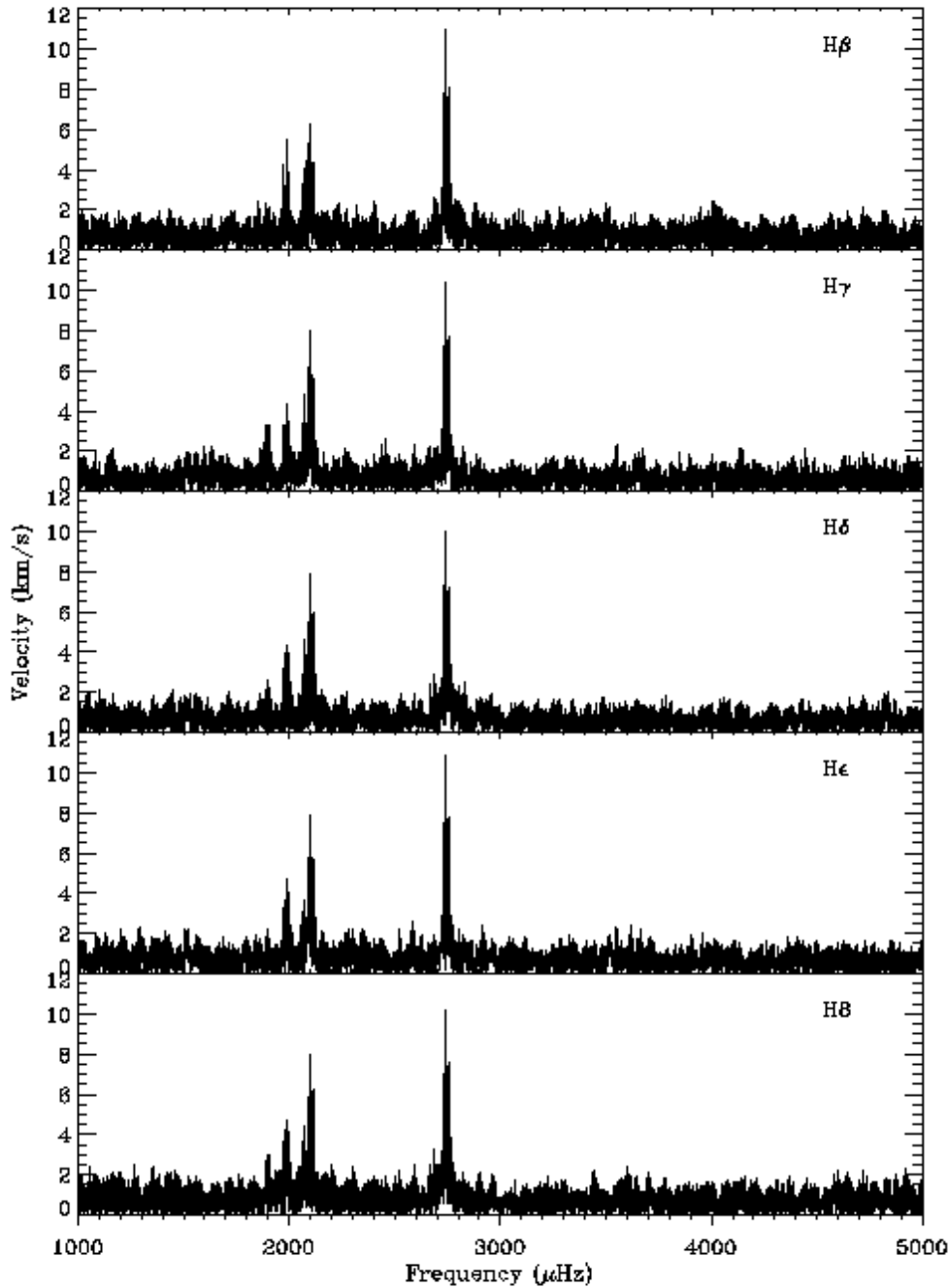,scale=0.75}

\caption{Amplitude spectra of five Balmer lines. No amplitude
  variation is evident in the dominant modes.}
\label{fig:may2000spec}
\end{figure*}

The velocity curve for H$\gamma$ is shown in Figure \ref{fig:velcurve}. It
has not been corrected for slow instrument drifts. The high quality of
the NOT data (3rd, 4th and 7th panels) -- due to its larger aperture --
suggests that fainter and/or shorter-period pulsating sdBs could be
observed with this telescope.

\section{Time-series Analysis}
\label{sec:tsa}
Because the quality of the observations varies through the data set, we
performed frequency analysis using a weighted Fourier Transform
\cite{K+F92}, where weights were assigned to each night in the May
2000 data set according to the internal scatter. Weights for the
March--April data were assigned to groups containing approximately
the same number of observations as one night in the May data.

\subsection{Velocities of Individual Balmer lines}
\label{sec:balmer}

The amplitude spectra of the five velocity time series from May 2000
are shown in Figure \ref{fig:may2000spec}. There does not appear to be any
substantial variation in velocity amplitude between the Balmer
lines. We have analysed all five time series, using the non-linear
least-squares multi-frequency fitting software \texttt{Period98}
\cite{p98}. The frequencies found are shown in Table
\ref{tab:may2000freq}. The same technique was applied to the data
presented in \pI\ (July 1999) and these frequencies are shown in
Table~\ref{tab:july1999freq}. White noise levels in the amplitude spectra
are typically $\sim$1000\,m\,s$^{-1}$ for each line in the July 1999
data, and 600--700\,m\,s$^{-1}$ in the May 2000 data.

The major difference between the two sets of frequencies is the
detection of the peaks at 1985\miHz\ and 2075\miHz\ in the May 2000
data set. In \pI\ it was uncertain whether the non-detection of the
latter (found to be the dominant mode in photometry by
\citebare{ECpaperX}) in any of the individual Balmer line time-series,
was due to variable amplitude or beating between closely spaced
modes. With a slightly longer time-series and reduced noise, this mode
is clearly detected in all Balmer lines. Somewhat surprising is the
detection of the mode at 1985\miHz, as this was found in photometry to
have an amplitude of only 3.3\,mmag, whereas the four other modes
detected had amplitudes greater than 13.9\,mmag. This mode may have
variable amplitude or may be a series of unresolved modes. There is
evidence for the latter possibility from \citeone{ECpaperX}, who found
three modes around 1985\miHz. As discussed further below, it is clear
that the crowded frequency spectrum of PG\,1605+072 severely
complicates the interpretation of the observations.

In \pI\ we reported the possible wavelength dependence of Balmer-line
velocity amplitudes (see Figure 3 of that paper). Here, we have tried
to confirm this result. Firstly, we found the weighted average frequencies
of each of the four modes that were detected in all
Balmer lines. These frequencies were then simultaneously fit to each
time-series, giving amplitudes and phases for each mode.

\begin{figure}
\epsfig{file=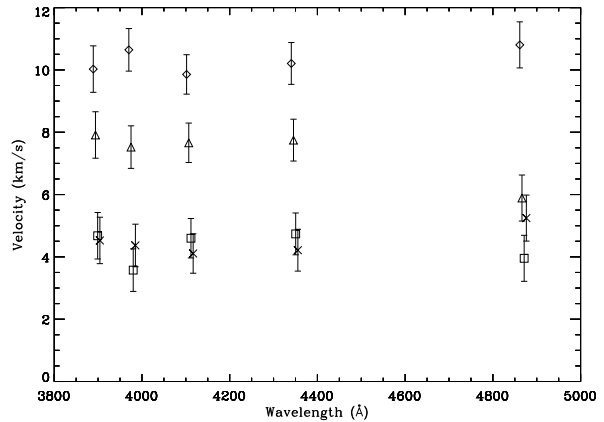,scale=0.47}
\caption{Velocity amplitude as a function of wavelength. Diamonds
represent the mode at 2742\miHz, triangles for 2102\miHz, squares for
2075\miHz\ and crosses for 1985\miHz. Small offsets to some points
have been added in the wavelength direction to improve clarity. There
is no clear dependence of amplitude on wavelength.}

\label{fig:ampwave2000}
\end{figure}

Figure \ref{fig:ampwave2000} shows Balmer-line velocity amplitudes as a
function of wavelength for these four modes. Error bars indicate the
white noise level in the 3000--5000\miHz\ region, where there appears
to be no excess power. For three of the modes, there is no clear
evidence for any wavelength dependence. For the 2102\miHz\ mode
(triangles), all lines except H$\beta$ appear to have the same
amplitude, with the H$\beta$ amplitude being slightly lower. We do not
feel this difference is significant enough to claim wavelength
dependence, particularly as this peak may in fact be a combination of
several closely spaced modes.

\begin{figure}
\epsfig{file=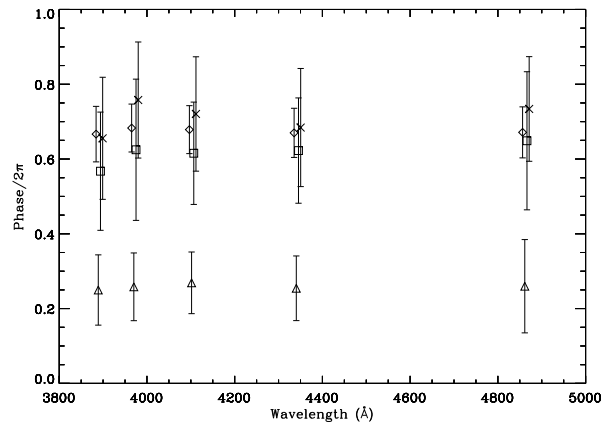,scale=0.47}
\caption{Oscillation phase as a function of wavelength. Symbols are
the same as in Figure \ref{fig:ampwave2000}. Small horizontal offsets have
been added for clarity.}
\label{fig:phiwave2000}
\end{figure}

\begin{table}
\caption{Frequencies (in $\mu$Hz) for five Balmer lines found from the
  May 2000 data.}
\begin{center}
\begin{tabular}{ccccc}
H$\beta$ & H$\gamma$ & H$\delta$ & H$\epsilon$ & H8 \\  \hline &
1891.26 &  &  & 1891.34  \\ 1985.73 & 1985.68 & 1985.58 & 1985.73 &
1985.79 \\ 2075.64 & 2075.72 & 2075.68 & 2075.68 & 2075.68 \\ 2102.21
& 2102.20 & 2102.17 & 2102.18 & 2102.24 \\ 2742.71 & 2742.68 & 2742.64
& 2742.70 & 2742.69 \\ \hline
\end{tabular}
\end{center}
\label{tab:may2000freq}
\end{table}

\begin{table}
\caption{Frequencies (in $\mu$Hz) for five Balmer lines found from the
  July 1999 data.}
\begin{center}
\begin{tabular}{ccccc}
H$\beta$ & H$\gamma$ & H$\delta$ & H$\epsilon$ & H8 \\  \hline &
1891.02 & 1891.00 & 1890.91 &  \\ 2102.14 & 2102.14 & 2102.11 &
2102.12 & 2102.12 \\ 2742.73 & 2742.47 & 2742.69 & 2742.67 & 2742.61
\\ \hline
\end{tabular}
\end{center}
\label{tab:july1999freq}
\end{table}

\subsection{Combined Time-series}
\label{sec:comb}

\begin{figure*}
\epsfig{file=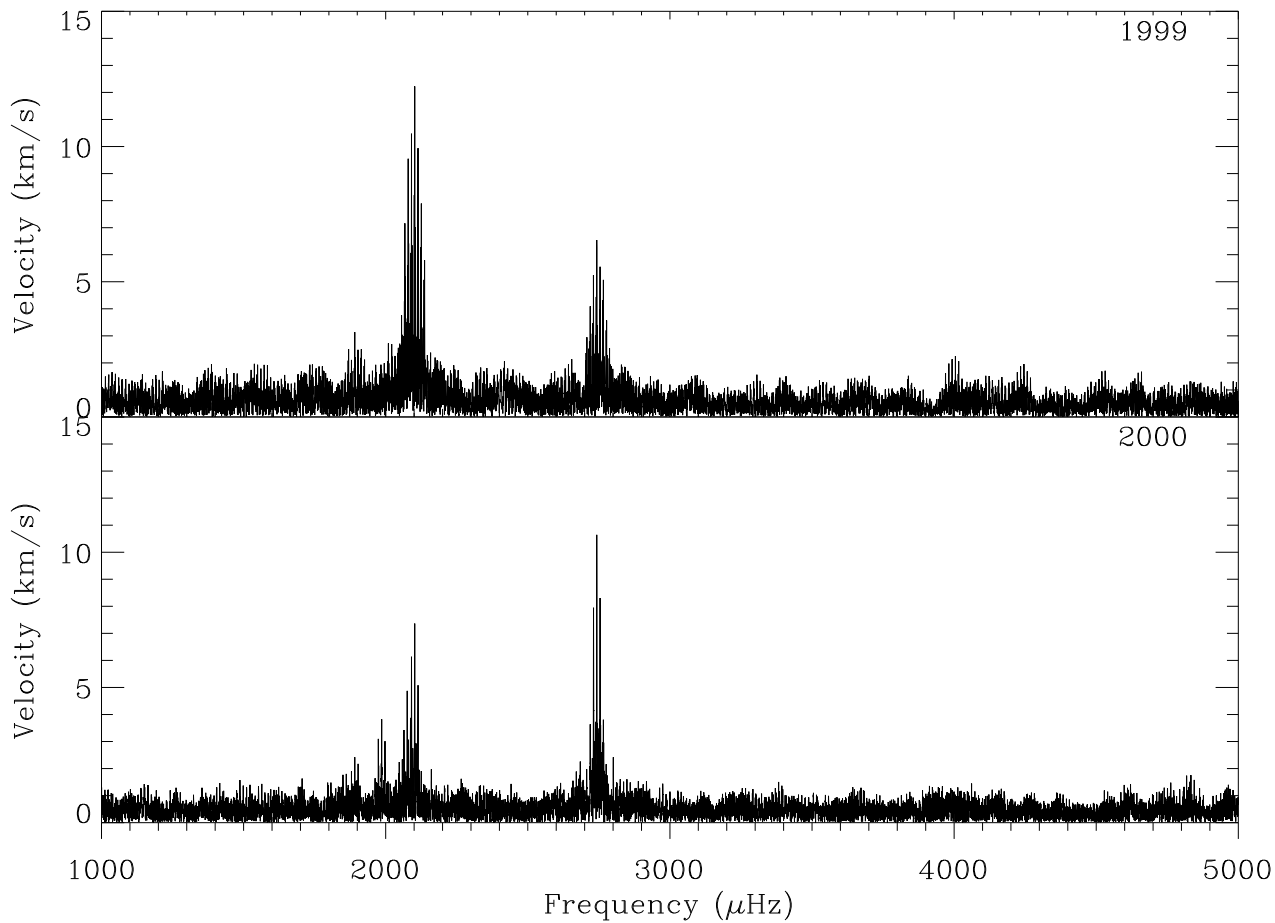}
\caption{Amplitude spectra for July 1999 \textit{(top)} and May 2000
\textit{(bottom)} based on a combined Balmer line time series. Note the
apparent change in amplitudes.}
\label{fig:2spec}
\end{figure*}

To reduce noise and perhaps detect more frequencies, we have combined
the time-series for the five Balmer lines. Before doing this, we
checked whether the oscillation phases were the same for each mode in
each time-series. This is simply a matter of examining the phases
found at fixed frequency. Relative phases are plotted as a function of
wavelength for each mode in Figure \ref{fig:phiwave2000}, with errors
determined using simple complex arithmetic (see
\citebare{BBV98a}). There is no evidence for phase variation with
wavelength, leaving us free to combine the time series.


To do this, we simply averaged the velocity at each observation. This
was possible since the weighting for each Balmer line was
approximately the same. Weights
were again derived from the internal scatter of each night. Combined
time series were constructed, for both July 1999 and May 2000. The two
resulting amplitude spectra are shown in Figure \ref{fig:2spec}. There
appears to be amplitude variation over the 10 month interval between
the observations, which we discuss in Section \ref{sec:sim}. The
white noise level is now 555\,m\,s$^{-1}$ for the July 1999 data set
and 425\,m\,s$^{-1}$ for the May 2000 data set, an improvement of
$\sim$45\% over the best single-line amplitude spectra.

The fitted frequencies are shown in Table \ref{tab:Hfreq2000}. The third
column contains the differences in frequency between the two sets of
data. The lengths of our two sets of observations are
$\sim7.9\times10^5$\,s (July 1999) and $\sim1.3\times10^6$\,s (May
2000). If the formal frequency resolution is given by $1/T$, then the
corresponding resolutions are $\sim$1.26\miHz\ and $\sim$0.77\miHz,
respectively. Comparing these values with those in the third column of
Table~\ref{tab:Hfreq2000}, we see that we have been able to estimate the
frequency to much better than the formal resolution. In the final
column we show the frequencies found from photometry, and again we
see agreement that is better than the formal resolution. This is not
surprising: at high signal-to-noise, the frequency of a strong peak in
the power spectrum can be estimated to a precision many times better
than the formal frequency resolution.



\begin{table}
\caption{Frequencies found after combining all Balmer lines in the
July 1999 and May 2000 data sets. Note that
$\Delta\nu=\nu_{2000}-\nu_{1999}$. The final column shows frequencies
found by \protect\citeone{ECpaperX} for comparison.}
\begin{center}
\begin{tabular}{cccc}
\hline $\nu_{2000}$ & $\nu_{1999}$ & $\Delta\nu$ & \citename{ECpaperX}
\\ ($\mu$Hz) & ($\mu$Hz) & ($\mu$Hz) & (\citeyear{ECpaperX}) \\ \hline
1891.19 & 1890.98 & $-$0.21 & 1891.42 \\ 1985.75 & --- & --- & 1985.32
\\ 2075.68 & 2075.29 & $-$0.39 & 2075.76 \\ 2102.21 & 2102.15 &
$-$0.06 & 2101.65 \\ 2742.71 & 2742.63 & $-$0.08 & 2742.72 \\ \hline
\end{tabular}
\end{center}
\label{tab:Hfreq2000}
\end{table}

\subsection{Extended Time-series}
\label{sec:ext}
\begin{figure}
  \epsfig{file=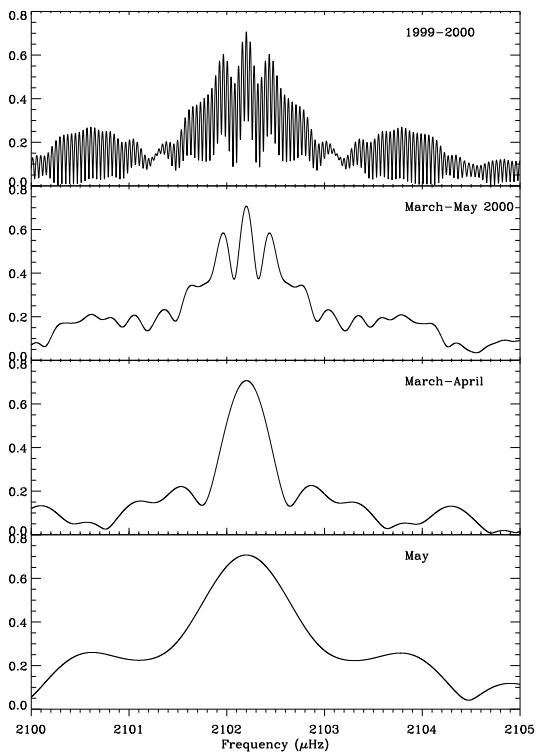,scale=0.43}
\caption{Spectral window (in amplitude) for 1999--2000, March--May
  2000, March--April and May data sets in the range
  2100-2105\miHz. Alias peaks are seen in the top two panels with a
  $\sim$0.25\miHz\ spacing, and in the top panel.}
\label{fig:zoomspecwin}
\end{figure}

To improve the frequency resolution still further, we now add the
data from March--April 2000 (Table \ref{tab:marapr-obs}) to the May 2000
series. For the extended time-series ($T\sim5.9\times10^6$\,s) the
formal frequency resolution is $\sim$0.17\miHz.  The longer time-series does
add complications however, since the number of alias peaks in the
amplitude spectrum increases significantly. There is a gap of 45
days between the midpoints of the two time series, leading to aliasing
that splits each peak into multiplets separated by
$\sim$0.25\miHz. The spectral window of each time series is shown in
Figure \ref{fig:zoomspecwin}. This includes the window function of all the
1999--2000 combined, which is discussed in Section
\ref{sec:allcombine}. The extra aliasing is evident in the top two
panels.

The frequencies and amplitudes determined from the extended
(March--May 2000) time-series
are shown in Table \ref{tab:ext-clean}. The white noise level for this
time-series is 370\,m\,s$^{-1}$, and we use this as an error for the
velocity amplitudes in column 3. The peaks at 2102 and 2742\miHz,
both shown as single in Tables \ref{tab:may2000freq} and
\ref{tab:july1999freq}, are now resolved into three and two peaks,
respectively. Also shown are the differences in frequency when
compared with photometry. If we take into account both our frequency
resolution and the ambiguity from alias peaks, the frequencies we find
are in excellent agreement with those found in photometry
\cite{ECpaperX}. In particular, the multiplicity of the peaks at 2102
and 2742\miHz\ agrees with the extended time series ($T\sim$33 days)
used by \citeone{ECpaperX}, who found 4 frequencies around 2102\miHz\
and 4 around 2742\miHz. They also found 5 frequencies around
2075\miHz, 6 around 2270\miHz, and 3 around 1985\miHz. We have not
detected these extra frequencies. It is not clear whether these
multiple peaks are real or whether they caused by amplitude
variability. We will for the moment consider them to be real, but will
address this question in more detail in Section \ref{sec:sim}.

In Table \ref{tab:ext-clean}, we note that the peaks at 2102.48\miHz\ and
2765.09\miHz\ are matched with quite low-amplitude modes found in
photometry ($\sim$1\,mmag), chosen because they are the closest in
frequency. The latter frequency can also be identified with a higher
amplitude mode detected by \citeone{ECpaperVII}. We note also that
2765.09\miHz\ is separated from $\sim$2742\miHz\ by twice the one
cycle per day alias frequency (i.e. 2$\times$11.57\miHz). The mode at
2742.47\miHz\ is also identified with a low amplitude mode, however it
was only found in the extended time series of
\citeone{ECpaperX}. Again, we have matched these modes purely because
they are the closest in frequency. This appears to give further
evidence for amplitude variability. 

\begin{table}
\caption{2000 frequencies, periods and amplitudes measured from the
extended time-series. The fourth column shows the amplitudes found
using the 2000 frequency solution, the fifth column shows the
difference between the frequencies found in this paper and those found
in photometry \protect\cite{ECpaperX}, and the sixth column shows
photometric amplitude rank.}

\begin{center}

\begin{tabular}{cccccc}
$\nu$ & $P$ & $V_{\mathrm{2000}}$ & $V_{\mathrm{1999}}$ 
& $\nu_{\mathrm{vel}}-\nu_{\mathrm{phot}}$ & Rank \\
($\mu$Hz) & (s) & (km/s) & (km/s) & ($\mu$Hz) & $n$ \\ 
\hline 
1891.01 & 528.82 & 1.99 & 3.18 & $-$0.41 & 5 \\ 
1985.75 & 503.59 & 4.13 & 0.63 & +0.43 & 8 \\ 
2075.72 & 481.76 & 4.27 & 2.11 & $-$0.04 & 1 \\ 
2101.57 & 475.83 & 3.40 & 4.97 & $-$0.08 & 3 \\ 
2102.48 & 475.63 & 8.47 & 11.1 & +0.04 & \llap{3}2 \\
2102.83 & 475.55 & 3.66 & 3.05 & $-$0.45 & 2 \\ 
2269.84 & 440.56 & 1.75 & 0.90 & $-$0.27 & 6 \\
2742.47\rlap{$^*$} & 364.63 & 4.45 & 3.57 & $-$0.25 & 4\rlap{$^*$} \\ 
2742.85 & 364.58 & 7.17 & 2.87 & +0.13 & 4 \\ 
2765.09 & 361.65 & 1.97 & 0.29 & $-$0.20 & \llap{2}1 \\ 
\hline
\end{tabular}
\end{center}
\label{tab:ext-clean}
$^*$This mode can also be identified as one of the extra frequencies
found in an extended time-series in Table 4 of \citeone{ECpaperX}.
\end{table}




To further quantify this variability, we have fit the frequencies
derived from the 2000 observations to our 1999 data. The amplitudes
from the fit are shown in the fourth column of Table
\ref{tab:ext-clean}, next to the amplitudes from the 2000 observations
for comparison. Errors in velocity are taken to be the white noise
level (555\,m\,s$^{-1}$). There are clearly large variations in most
peaks (even more so, considering that the weakest 1999 amplitudes
should be considered as upper limits).

\begin{figure}
\epsfig{file=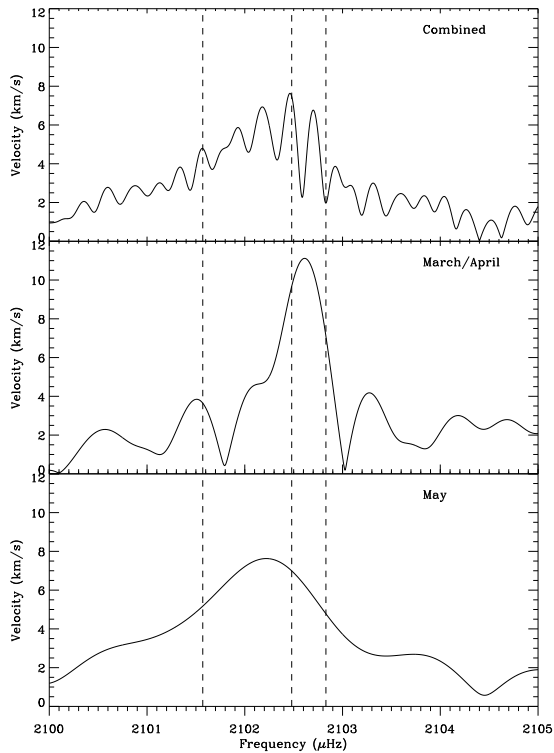,scale=0.43}
\caption{Amplitude spectra for the Combined, March/April and May data
  sets in range 2100--2105\miHz. The dashed lines indicated
  frequencies measured from the extended time series.}
\label{fig:zoomspec2100}
\end{figure}

\begin{figure}
\epsfig{file=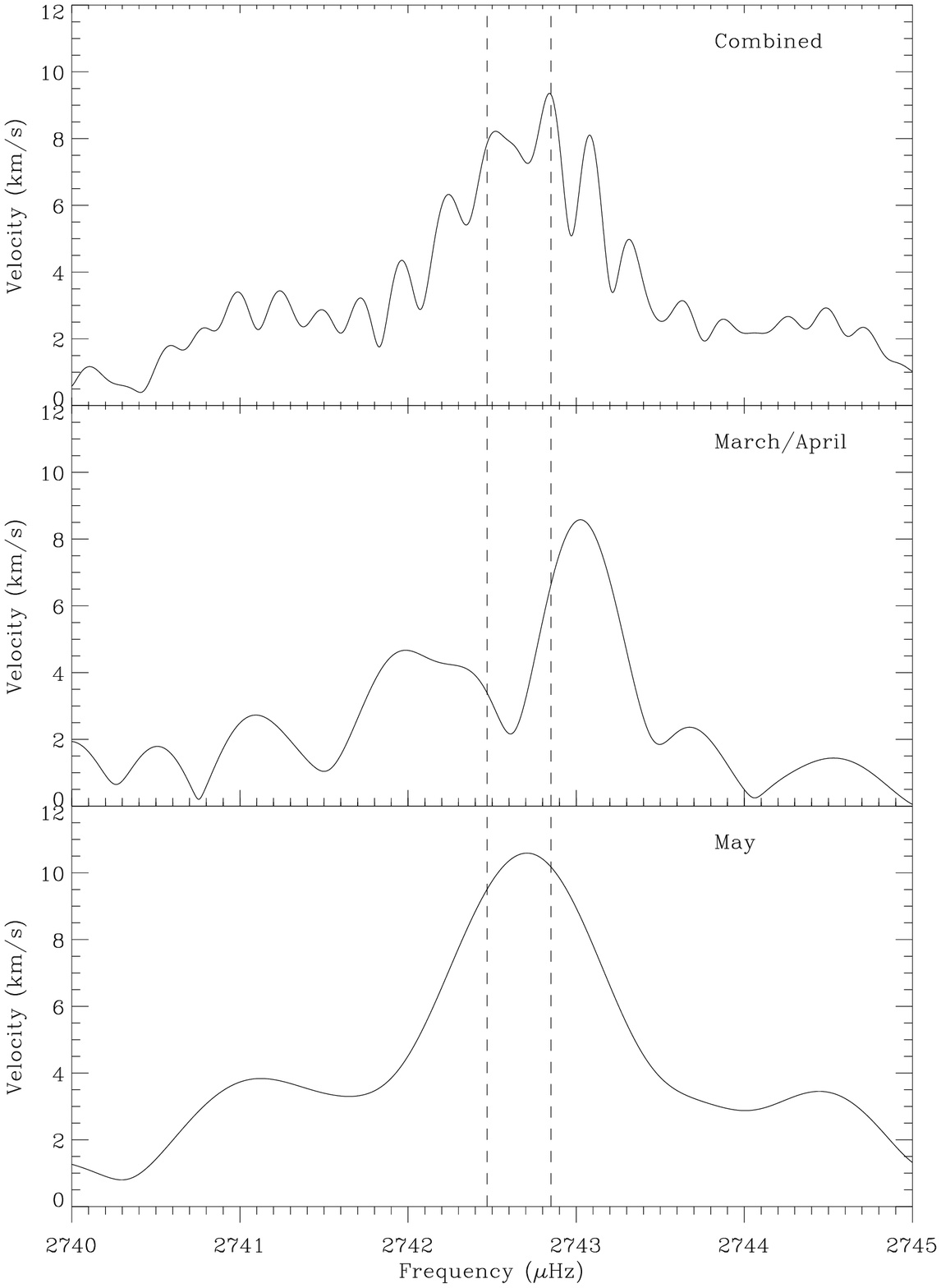,scale=0.43}
\caption{Same as Figure \protect\ref{fig:zoomspec2100} expect in the range
2740--2745\miHz.}
\label{fig:zoomspec2740}
\end{figure}

\begin{figure}
\epsfig{file=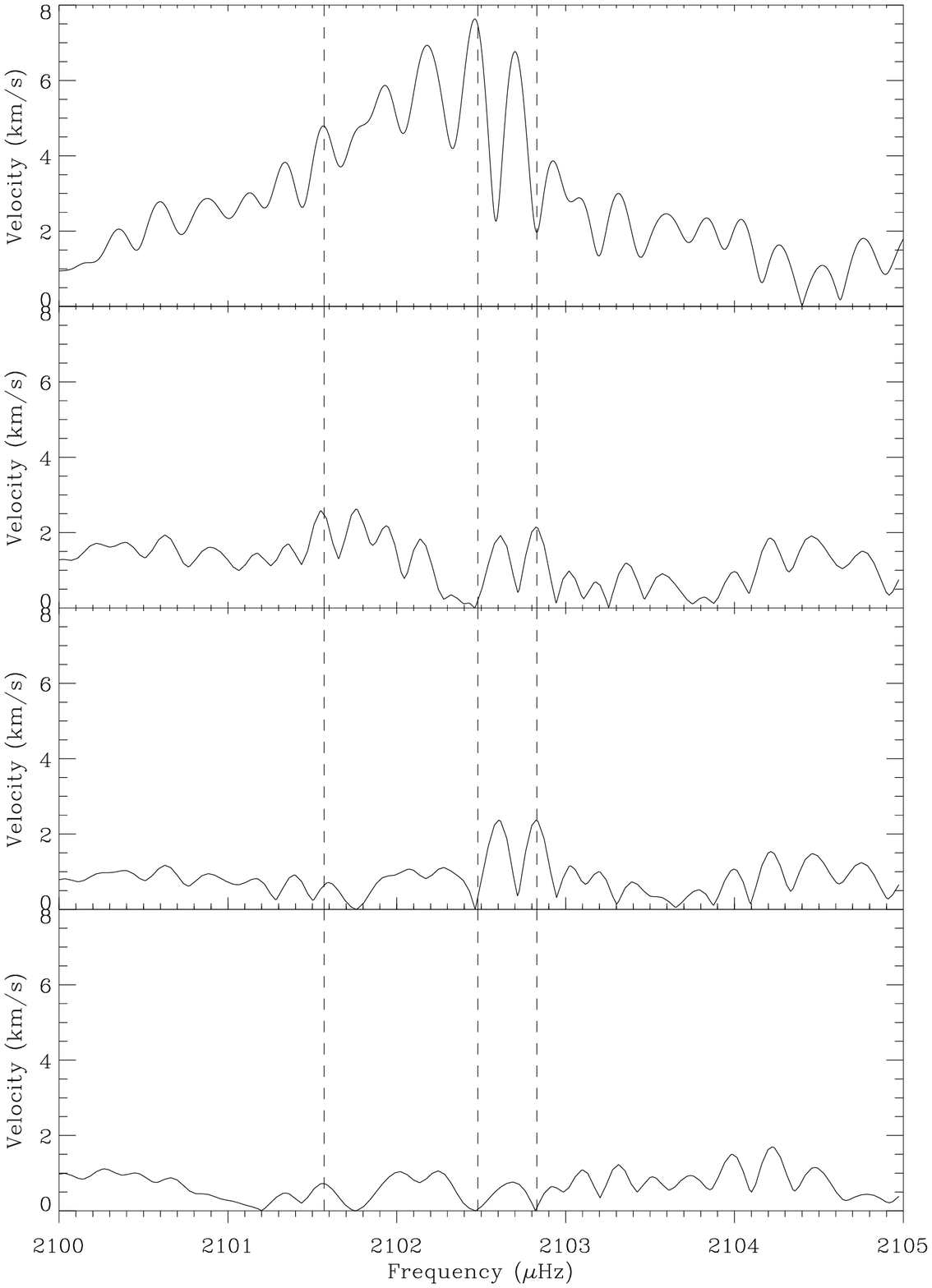,scale=0.43}
\caption{Amplitude spectra of the 2100--2105\miHz\ region showing the
  successive removal of 3 closely spaced frequencies.}
\label{fig:clean2100}
\end{figure}


\section{Simulations}
\label{sec:sim}
How well resolved are the closely spaced multiplets at 2102\miHz\ and
2742\miHz\ shown in Table \ref{tab:ext-clean}? To answer this, we show the
regions in question in Figures \ref{fig:zoomspec2100} and
\ref{fig:zoomspec2740}. From the top panel of Figure \ref{fig:zoomspec2100} it
is clear that the frequency resolution is sufficient to resolve the power at
2102\miHz\ into at least two peaks, perhaps three. The two frequencies
found around 2742\miHz\ also appear to be resolved in
Figure~\ref{fig:zoomspec2740}. In the bottom panel of Figure \ref{fig:2spec} it
appears that the peak around 2742\miHz\ is higher than the one around
2102\miHz, however Table \ref{tab:ext-clean} shows the opposite. This
can be explained by examining the amplitude spectra (shown in
Figure~\ref{fig:clean2100}) around 2102\miHz\ after each frequency is
removed. It is still not clear, however, whether these frequencies are
real or an artifact of the variation in amplitude of one frequency.

\begin{figure}
  \epsfig{file=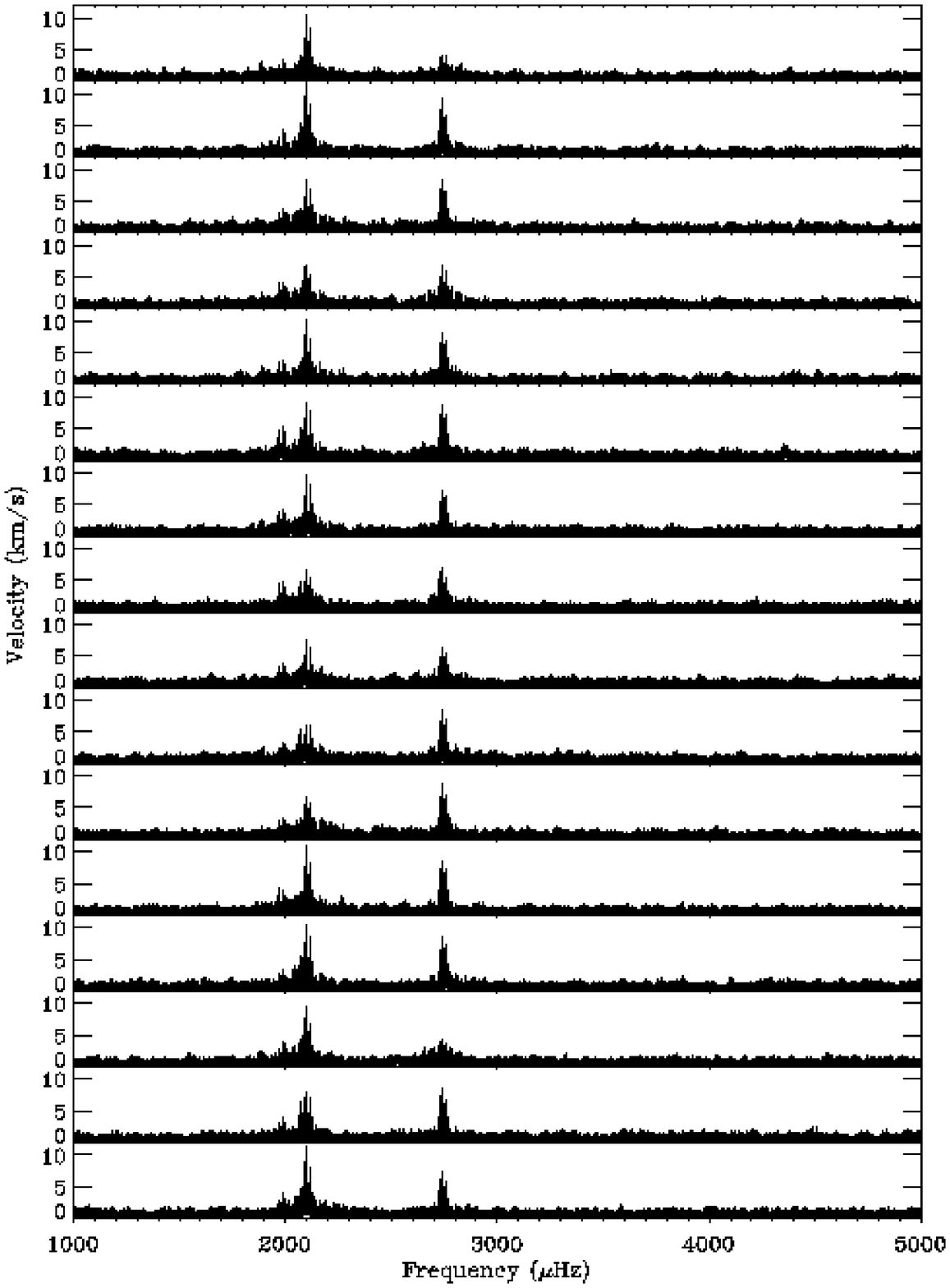,scale=0.47}
  \caption{Simulated amplitude spectra using the 10 frequencies from
    Table \protect\ref{tab:ext-clean} as input. These spectra are based on
    the window function of all observations from 2000. It is clear that
    the variable amplitude of the highest peaks is due to beating.}
  \label{fig:simplot}
\end{figure}

To investigate these effects further, we have created simulated time
series for both our sets of observations (1999 and 2000). The inputs
are the 10 frequencies measured from the time-series with the highest
frequency resolution (i.e, 2000) with their corresponding
amplitudes. Note that each
simulation had exactly the same input amplitudes and frequencies, and
only the phases were randomised. We have used two different noise
levels, $\sim$600\,m\,s$^{-1}$ and $\sim$300\,m\,s$^{-1}$, as these
bracket the noise in our observations. The same phases were used at
both noise levels. The basic features of the two
different groups of spectra are the same, implying that noise does not
play a significant role in the shape of the amplitude spectrum. A
selection of simulated amplitude spectra with noise
$\sim$600\,m\,s$^{-1}$ are shown in Figure \ref{fig:simplot} using the
2000 window function. 

The most
interesting feature of the majority of these
spectra is the variability of the amplitudes of the peaks at
$\sim$2102\,$\mu$Hz and at $\sim$\,2742\,$\mu$Hz. This observed
variability does not depend strongly on the window function and arises
from beating between the closely spaced modes around these two
frequencies.  However in Table \ref{tab:Hfreq2000}
there is only a small difference ($<$0.1\miHz) between frequencies
measured in July 1999 and May 2000. If there were two or more
frequencies beating, we would expect the peaks to move by more
than this difference, since the measured frequencies span a range of
$\sim$1.3\miHz. 



\citeone{ECpaperX} found 2 frequencies around 2102\miHz\ in both
halves of their $\sim$15 night time series, as well as in the full
series. They also found weak ``satellite'' frequencies near almost
all of the highest amplitude peaks. %
%
%
We find only one peak at 2102\miHz\ in our May 2000
observations, however weighting the data may degrade the frequency
resolution. The middle panel of Figure \ref{fig:zoomspec2100} suggests
that 2 peaks may be clearly resolved in the March/April 2000
observations. %
%
%
If we assume there are 3 frequencies around 2102\miHz, we detect
2 frequencies in $\sim$90\% of our simulations. It seems that our
detection of only one peak may be due simply to chance. Around
2742\miHz\ only one frequency is always resolved, with the 2nd
frequency resolved $\sim$50\% of the time. This suggests a minimum of
two frequencies around 2102\miHz, and possibly two around
2742\miHz. Two frequencies around 2742\miHz is supported by
\citeone{MR_PhD02}, who finds a similar separation ($\sim$0.3\miHz) in
a re-analysis of the \citeone{ECpaperX} observations. By combining all
of our data (1999--2000), we are in a position to make one final test.





\section{Combination of all observations}
\label{sec:allcombine}

\begin{figure}
  \begin{center}
    \leavevmode
    \epsfig{file=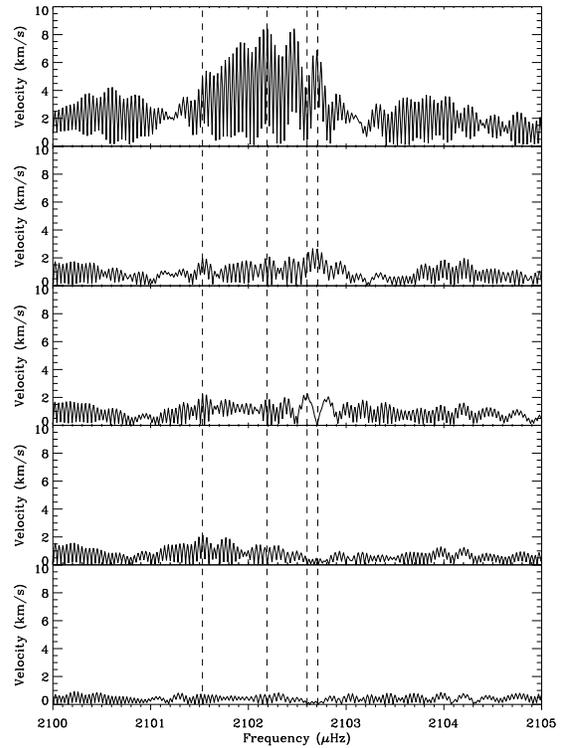,scale=0.43}
    \caption{Same for Figure \ref{fig:clean2100}, except with combined
    1999 and 2000 time series.}
    \label{fig:allcomb}
  \end{center}
\end{figure}

If we combine all of our observations (1999--2000), we find that the
amplitudes of each peak is approximately the same
($\sim$8\,km\,s$^{-1}$). This implies that the oscillations are
coherent over this one year period, and we can therefore perform a
frequency analysis on the time series.
Combining all of the data
introduces fine structure in the spectral window, as shown in
the top panel of Figure \ref{fig:zoomspecwin} in the region around
2100\miHz. The frequencies we measure from
this time series are the same as those found in the March--May 2000
data, except around 2102.5\miHz, where we now find 3 frequencies instead
of two. These frequencies, while they describe the observations (see
Figure \ref{fig:allcomb}), are split such that we consider them to be
caused by amplitude variation. The two frequencies not found in
1999--2000, but found in the March--May 2000 data (2269.84 and
2765.09\miHz), we consider marginal detections. 
We can now conclude that the frequencies in Table \ref{tab:realfreq},
derived from the March--May 2000 observations, are real.

\begin{table}
\caption{Detected frequencies and periods in velocities of
  PG\,1605+072. These are based on the March--May 2000 observations,
  but also found in a time series of all observations (1999--2000).}
\label{tab:realfreq}
\begin{center}
\begin{tabular}{cc}
Frequency ($\mu$Hz) & Period (s) \\
\hline
1891.01 & 528.82 \\ 
1985.75 & 503.59 \\
2075.72 & 481.76 \\
2101.57 & 475.83 \\
2102.48\rlap{$^*$} & 475.63 \\
2742.47 & 364.63 \\
2742.85 & 364.58 \\
\hline
\end{tabular}
\end{center}
$^*$We include this frequency as representative of a mode with
variable amplitude around 2102.5\miHz.
\end{table}

\section{Comparison with Woolf et al. (2002)}
\label{sec:comp}
Our observations were by coincidence concurrent with those of
\citeone{WJP01}, so we are in a position to examine some of their
findings. Firstly, based on data taken over 16.3 hours across a 32.1
hour period, they found an apparent power shifting between modes, which
we also find, and attribute to both amplitude variability over time
and beating between closely spaced frequencies. They
also found evidence for rotational splitting of the 2742\miHz\ mode,
and derived a rotational period of 12.6$\pm$2.8 hours. This is not in
agreement with \citeone{HRW99}, who derived an upper limit on the
rotational period of 8.7 hours from spectral analysis. While the
pulsations cause some broadening of the spectral lines, the pulsation
amplitudes are not high enough to shift \citename{HRW99}'s upper limit
into the range found by \citename{WJP01} The rotational splitting they
found is $\sim$11\miHz, which is very close to one cycle per
day. Because of this, the reality of this splitting should probably be
treated with scepticism.

\begin{figure}
\epsfig{file=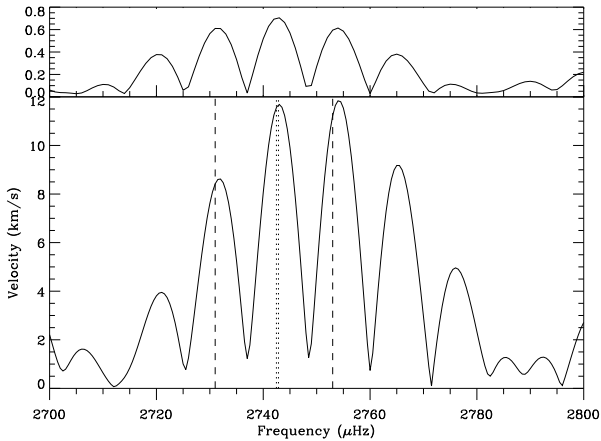,scale=0.47}
\caption{Amplitude spectrum of nights MJD 51675.1--51676.4
  (\textit{bottom}), showing no rotational splitting of the 2742\miHz\
  mode. The frequencies found by \protect\citeone{WJP01} are plotted
  as dashed lines. The peaks they correspond to are alias peaks. For
  comparison, the frequencies found in this work are plotted as dotted
  lines. The window function for the two nights (\textit{top}) is also
  shown.}

\label{fig:nosplit}
\end{figure}

Some of our observations overlap with \citename{WJP01}'s, and we have
plotted in Figure \ref{fig:nosplit} the amplitude spectrum of the two
nights closest in time to theirs. Our observations for this period start
at MJD 51675.1 and finish at MJD 51676.4, while the \citename{WJP01} start
at MJD 51675.9 and finish at MJD 51677.25. The dashed lines in the figure
indicate the frequencies found by Woolf et al, which appear to be
one cycle per day alias peaks. The dotted lines indicate the
frequencies found in this paper. For comparison we have shown the
spectral window of our observations in the top panel. The dashed lines
correspond very well to the alias peaks in our spectral window, and we
conclude that this is most likely what \citename{WJP01} have
identified, although the central peak (which we have identified with
the frequency found in photometry) is not apparent in their amplitude
spectra. We feel timing errors may be a possible explanation for the
splitting, and lack of this central peak.

\section{Discussion}
\label{sec:disc}
The model of \citeone{Kawaler99} is the best pulsation model for
PG\,1605+072 produced so far, although it only matches 5 of the dominant
frequencies from \citeone{ECpaperX} (they detected up to 55
frequencies, although some may be artifacts of amplitude
variation). \citename{Kawaler99}'s model predicts an equatorial
velocity of $\sim$130\,km\,s$^{-1}$ leading to a rotational period of
around 3 hours, in agreement with the upper limit found by
\citeone{HRW99} of $P_{\mathrm{rot}}<8.7\mathrm{h}$
($v$sin$i=39$\,km\,s$^{-1}$).


What are the possible reasons for modes which are so closely
spaced? Rotation has already been mentioned, however the minimum
rotational splitting for an $l=1$ mode is $\sim$\,16\,$\mu$Hz to first
order, and for an $l=2$ mode is $\sim$\,27\,$\mu$Hz. This is based on
the maximum rotational period of \citeone{HRW99}, and assumes that the
star oscillates in $g$-modes \cite{Kawaler99} in a similar way to a
white dwarf. This is not the case, however, since unlike white dwarfs,
horizontal motion does not dominate vertical motion on the surface of
sdBs; in fact, they are of the same order (Kawaler, private
communication). Despite this, the splitting values above are minima,
so it is clear rotation is not the only effect causing splitting.
Strong magnetic fields also cause splitting of oscillation
modes, however little work has been done to measure fields of
sdBs. Finally, we feel that amplitude variability over several months
can explain at least some of the observed splittings. %
More
detailed modelling of PG\,1605+072 using the so-called second
generation pulsating sdB models of \citeone{CFB01} may help to answer
these questions.

\section{Conclusions}
\label{sec:conc}
Of all the pulsating sdBs currently known, PG\,1605+072 is clearly one
of the most interesting. Using our newest observations we make several
conclusions. 

\begin{itemize}

\item From our main campaign we have detected 7 oscillation
  frequencies from velocity variations in the star. These frequencies
  are shown in Table \ref{tab:realfreq}. Most
  of these correspond to the strongest oscillation frequencies found
  in photometry.

\item We find the velocity amplitudes and phases in the individual
  Balmer lines to be equal and cannot confirm the wavelength
  dependence of amplitude seen in \pI.

\item Apparent amplitude variability between our two sets of observations
  can be explained by a combination of beating between closely spaced
  modes and amplitude variability.

\item To fully resolve and identify as many closely spaced frequencies
  as possible, it is necessary to have both good time coverage to
  reduce aliasing effects, and most importantly for PG\,1605+072, to
  have a time series that is as long as possible, thus improving the
  frequency resolution in the amplitude spectrum.

\item We can explain the possible rotational splitting of the
  2742\miHz\ peak claimed by \citeone{WJP01} as an aliasing effect
  (possibly caused by incorrect time stamps), since our
  observations are concurrent.
\end{itemize}

PG\,1605+072 offers asteroseismologists several challenges. It has many
oscillation frequencies, although some have variable amplitudes;
it has longer periods than most other pulsating sdBs, however it is
rotating fast enough that rotationally split modes may be unequally
spaced; it may be oscillating in $p$ or $g$-modes or a combination of
both. Further theoretical work is needed in these areas.
We feel that despite these challenges, a
campaign of simultaneous photometry and spectroscopy will
help to answer the questions we have raised here, and perhaps
allow us to comprehend one of the least understood phases of stellar
evolution.


The authors would like to thank Steve Kawaler for helpful discussions.
This work was supported by an Australian Postgraduate Award (SJOT),
the Australian Research Council, the Danish National Science
Research Council through its Center for Ground-based Observational
Astronomy, and the Danish National Research Foundation through its
establishment of the Theoretical Astrophysics Center.


\end{document}